%Paper: hep-th/9301014
%From: "Don Spector" <das@hepth.cornell.edu>
%Date: Tue, 5 Jan 93 15:32:27 EST
%Date (revised): Mon, 19 Jun 1995 14:30:49 -0400

%
\input harvmac.tex
%\draftmode
\pretolerance=10000
\def\B{Bogomol'nyi }
\def\PS{Prasad-Sommerfield }
\def\Stilde{{\tilde S}}

%---------------------------
\Title{HWS/92-07, CSULB-HEP-45}{
\vbox{\centerline{Solitons and Instantons with(out)
Supersymmetry}}}

\ifx\answ\bigans{

\centerline{Zvonimir Hlousek\footnote{*}{hlousek@beach.csulb.edu} }
\medskip\centerline{Department of Physics and Astronomy}
\centerline{California State University, Long Beach}
\centerline{Long Beach, CA \ 90840 USA}
\vskip.2in
\centerline{Donald Spector\footnote{$^\dagger$}{spector@hws.bitnet} }
\medskip\centerline{Department of Physics, Eaton Hall}
\centerline{Hobart and William Smith Colleges}
\centerline{Geneva, NY \ 14456 USA}
}
\else {

\centerline{\qquad\qquad Zvonimir
Hlousek\footnote{*}{hlousek@beach.csulb.edu}
\hfill Donald Spector\footnote{$^\dagger$}
{spector@hws.bitnet}\qquad\qquad}
\bigskip\centerline{\qquad\qquad Department of Physics and
Astronomy
\hfill Department of Physics, Eaton Hall\qquad\qquad }
\centerline{\qquad\qquad California State University, Long Beach
\hfill Hobart and William Smith Colleges\qquad\qquad }
\centerline{\qquad\qquad Long Beach, CA 90840 USA \hfill Geneva, NY
14456 USA\qquad\qquad }
}
\fi

\vskip .3in
We give model-independent arguments, valid in nearly any number of
spacetime dimensions, that topological solitons and
instantons satisfy Bogomol'nyi-type bounds and, when these bounds
are saturated, satisfy self-duality equations. In the supersymmetric
case, we
also show that, in spacetime dimensions greater than two, theories with
topological charges necessarily exhibit extended supersymmetry, in which
the
topological charge appears as the central charge. The significance
of our arguments lies in their
generality. In the supersymmetric case, we obtain insight into
the contrast observed between topological charges in 1+1 and higher
dimensional models.
The centerpiece of our method is to require that the supersymmetric
extension of a generic (non-supersymmetric) field theory be
self-consistent.

\Date{11/92; revised 4/94}

\newsec{Introduction and Summary}

The topological classification of field configurations is an
important ingredient in our non-perturbative
understanding of field theories. Topological solitons and
instantons have, in particular, received a great deal of
attention.
The basic theory of these field configurations is covered
well in \ref\Raj{R. Rajaraman, {\it Solitons
and Instantons}, Amsterdam, North-Holland,
%Physics Publishing,
1987.},
which also contains a comprehensive list of references to the
original literature.

The study of solitons and instantons in various
models has revealed important similarities between these two types
of field configurations. Solitons generally exhibit \B bounds (their
energy is bounded from below by the magnitude of the topological
charge); instantons generally exhibit analogous bounds, with their
(Euclidean) action bounded from below by the magnitude of the
instanton number.
When field configurations saturate such a bound, they not only
automatically satisfy the equations of motion, but they also, in
fact, generally satisfy simpler first-order equations, called
\B or self-duality equations. Perhaps the most
familiar examples are the \B equations
\ref\Bog{E.B. Bogomol'nyi, {\it Sov. J. Nucl. Phys.}
{\bf 24} (1976) 449.} for the 't Hooft-Polyakov
monopole \ref\mono{G. 't Hooft, {\it Phys. Rev.} {\bf D14} (1976)
3432\semi
A.M. Polyakov, {\it JETP Lett.} {\bf 20} (1974) 194.}
in the \PS limit \ref\PS{M.K. Prasad and C.H.
Sommerfield, {\it Phys. Rev. Lett.} {\bf 35} (1975) 760.},
and the self-duality equations \ref\selfd{A.A. Belavin, A.M.
Polyakov, A.S. Schwartz, and Yu.S. Tyupkin, {\it Phys. Lett.}
{\bf 59B} (1975) 85.}
for Yang-Mills instantons \ref\inst{G. 't Hooft, {\it Phys. Rev.
Lett.} {\bf 37} (1976) 8\semi
G. 't Hooft, {\it Phys. Rev.} {\bf D14} (1976) 3432\semi
A.M. Polyakov, {\it Nucl. Phys.} {\bf B121} (1977) 429.}
in {\hbox{$3+1$}} dimensions.
We will use the phrase ``\B relationships'' to refer
collectively to \B bounds and \B equations.

In recent work, we began to develop a model-independent approach to
demonstrating the appearance of \B relationships \ref\letters{Z. Hlousek
and
D. Spector,
{\it Mod. Phys. Lett.} {\bf A7} (1992) 3403\semi
D. Spector, {\it Bogomol'nyi Explained: An Application of N=0
Supersymmetry},  in {\it Particle Physics from Underground to
Heaven: the Proceedings of the 15th Johns Hopkins Workshop on
Current Problems in Particle Theory}, World Scientific, 1992).}
\ref\extsusy{Z. Hlousek and D. Spector, {\it Nucl. Phys.} {\bf
B370} (1992) 143.}
\ref\bogex{Z. Hlousek and D. Spector, {\it \B Explained},
{\it Nucl. Phys.} {\bf B397} (1993) 173.}.
That work suffered from two limitations: it was valid
only in $2+1$ and $3+1$
dimensions, and it pertained only to solitons, not to instantons.
In this paper, we address those shortcomings,
extending our arguments to solitons in any number of spacetime
dimensions, $2+1$ or higher; and to instantons in any number of
Euclidean dimensions, two or higher.
Our method here, as in our previous papers, is to require
that the supersymmetric extension of a theory behave self-consistently.
Thus, here, as an essential step, we extend our earlier work
in the supersymmetric case to higher
dimensions as well, showing that in $2+1$ or more dimensions, any theory
with
$N=1$ supersymmetry and a conserved topological charge automatically has
extended supersymmetry in which the topological charge appears as the
central
charge.

Throughout this paper, since we are only examining topologically
non-trivial soliton and instanton field configurations,
we will, for simplicity, use the term ``soliton''
to mean any field configuration
with a non-zero value of a conserved topological charge (given by an
integral
over all space), and the term ``instanton''
to mean any Euclidean field configuration with a
non-zero value of a topological index (given by an integral over all
spacetime).

Note that the importance of our work is
that it provides a {\it general} explanation for properties
of solitons and instantons which appear routinely but which
have been understood only in a model-dependent way. Consequently, we
focus here on presenting those general arguments.
For specific examples, we will refer the reader to the literature, which
already contains analyses of a number of models.
Furthermore, since this work generalizes our earlier results, we will
concentrate here on what is new,
and simply summarize our earlier results as needed.
The reader interested in the details of our earlier
work should consult our papers listed in the references.

In the interests of clarity, we will divide the argument into two
pieces, one supersymmetric and one not.
In the next section, we will show that in any theory in $2+1$ or
more spacetime dimensions with simple supersymmetry and a
topological charge, an extended superalgebra appears, in which
additionally the topological charge appears as the central charge
of the extended superalgebra.  Then, in
the subsequent section, we will use this result as the basis for
arguments that solitons and instantons in general (non-supersymmetric)
theories exhibit \B bounds.

Finally, the reader is reminded that \B relationships are classical
statements.  The extended superalgebra we uncover in the next section
does appear in both the classical and quantum theories.
The other results in the paper, however, which depend, for example,
on the equations of motion, are to be understood as classical results
only.

\newsec{The Argument, Part 1: Solitons and Extended
Supersymmetry}

In \ref\OW{D. Olive and E. Witten, {\it Phys. Lett.} {\bf 78B} (1978)
97.}, Olive and
Witten discovered that in a number of models with
supersymmetry and a
topologically conserved charge, the full invariance algebra is
actually an
extended supersymmetry algebra in which the topological charge
appears as the
central charge. This type of structure was found again
in Chern-Simons models with vortices \ref\LLW{C. Lee,
K. Lee, and E. J. Weinberg, {\it Phys. Lett.} {\bf 243B} (1990)
105.},
and in the $2+1$
dimensional $O(3)$ non-linear sigma
model \ref\ZHDS{Z. Hlousek and D. Spector, {\it Nucl. Phys.} {\bf
B344} (1990) 763\semi
see also
 E. Witten, {\it Phys.Rev.} {\bf D10} (1977) 2991, and
the first item in \letters.}, and in 10-dimensional string
and string-related models \ref\tendim{J.A. de Azcarraga, J.P. Gauntlett,
J.M. Izquierdo, and P.K. Townsend, {\it Phys. Rev. Lett.}
{\bf 63} (1989) 2443\semi M. Cvetic, F. Quevedo, and S.-J. Rey,
{\it Phys. Rev. Lett.} {\bf 67} (1991) 1836.}.
In \extsusy,
a general explanation for this phenomenon was obtained for $2+1$
and $3+1$
dimensional models. Here we extend that work to an arbitrary
number
of spacetime  dimensions (greater than two), and we see explicitly
why the case of $1+1$ spacetime dimensions is different.

Consider a theory with $N=1$ supersymmetry and a conserved
topological charge.
When a charge is conserved topologically, this means that the
conservation law
holds without use of the equations of motion.
Consequently the conserved current $J_\mu$ can be
written as the curl of a potential \ref\pot{C. Cronstr\"om and J.
Mickelsson, {\it J. Math. Phys.} {\bf 24} (1983) 2528\semi
R. Jackiw {\it in}  S. Treiman, R. Jackiw, B. Zumino, and E.
Witten {\it Current Algebra and Anomalies} (World Scientific,
Singapore, 1985).}.
In $d>2$ spacetime dimensions, this means
\eqn\curleqn{J_{\mu_1} = \epsilon_{\mu_1 \cdots
\mu_d}\partial^{\mu_2}
   A^{\mu_3 \cdots \mu_d}~.}
The potential $A$ is a $(d-2)$-index antisymmetric tensor. (We find
it convenient to use tensors rather than forms.)
This potential is not a fundamental field, but is an expression
defined in terms of the fields of the theory.
Now the potential is not uniquely defined; rather, it is only
defined up to
gauge transformations. Under
\eqn\Atransf{A^{\mu_1 \cdots \mu_{d-2}}\rightarrow
   A^{\mu_1 \cdots \mu_{d-2}} +
   \partial^{[ \mu_1} \Omega^{\mu_2\cdots\mu_{d-2}]}}
the current $J_\mu$ is unchanged.
As we show in the appendix, there is a choice of
gauge in which $A$ is divergenceless.
Hereon, we use that gauge exclusively.

Now recall that we are considering a theory with $N=1$
supersymmetry.
We denote
the supercharges $Q_\alpha$, where $\alpha$ is a spinor index.
Define
\eqn\secsusy{\Stilde^\alpha_{\mu_1} =
[Q^\beta,A_{\mu_1\cdots\mu_{d-2}}]
   (\gamma^{\mu_2}\cdots\gamma^{\mu_{d-2}})^\alpha_\beta ~.}
(Note that this expression is only sensible for $d\ge 3$.)
Since the
supersymmetry charges commute with the translation generators,
and $A$ is divergenceless,
we have that $\Stilde$ is a conserved spinor current.
Furthermore, the supersymmetry
transform of $\Stilde$ contains the topological current.
This is easy to see, as the topological current is the
field strength associated with the potential, and
so appears at two orders of $\theta$ above the potential.
Explicitly, applying
two supersymmetry transformations produces a momentum
generator,
and so \eqnn\centchg
$$\eqalignno{
\{Q_\alpha,\Stilde^\alpha_{\mu_1}\}
  &= \partial^\beta_\alpha A_{\mu_1\cdots\mu_{d-2}}
    (\gamma^{\mu_2}\cdots\gamma^{\mu_{d-2}})_\alpha^\beta
      + \cdots \cr
  &\supset \epsilon_{\mu_1 \cdots \mu_d}\partial^{\mu_2}
   A^{\mu_3 \cdots \mu_d} = J_{\mu_1} ~, &\centchg \cr}$$
and thus the topological current appears in the Lorentz decomposition
of $\{Q_\alpha,\Stilde^\alpha_\mu\}$.
Since the $N=1$ supersymmetry transform of $\Stilde$ contains
the topological current,
then $\Stilde$ must not only be non-trivial but must be distinct
from the
original supercurrent \ref\HLS{R. Haag,
J. {\L}opusza\'nski, and M. Sohnius, {\it Nucl. Phys.} {\bf B288}
(1975) 257.}.
Thus the charge associated with $\Stilde$ is a new
conserved spinor charge which
transforms into the topological charge under the
original supersymmetry, and so this theory has an
$N=2$ supersymmetry invariance in which the topological charge
appears as
a central charge. This completes the argument, and we have shown what
we wished to show in this section.

We make four comments to conclude this section.
First, there may be other non-trivial terms in the $N=1$ supersymmetry
transform of $\Stilde$; this in fact generically
occurs in even spacetime dimensions where one can have scalar
and pseudoscalar central charges \OW \ref\even{Z. Hlousek and
D. Spector, {\it Phys. Lett.} {\bf 283B} (1992) 75.}.
In \centchg, we have explicitly identified only the terms where the
topological charge appears.
Second, note that we are not asserting that
there is a canonical $N=2$ field content, but rather that with the
$N=1$ field
content, there is already a second supersymmetry (as happens in
various of the
specific models people have studied). Third, it is possible that
our construction will produce a non-local $\Stilde_\mu^\alpha$
(stemming
at heart from the possibility of a non-local solution to (A.2)
(see the appendix)).
This is not significant.
The algebraic constraints on a theory are of course every
bit as real whether they come from a conserved charge associated with
a local or non-local conserved current.  Furthermore, if a non-local
solution does occur, this should merely reflect the absence
of a sufficient number of auxiliary fields.
And fourth and finally, notice that our index
manipulations explicitly break down in $1+1$ dimensions.
This makes sense. The potentials for topological currents in $1+1$
dimensions are scalars, and so there is no gauge transformation for such
a potential, no way to require such a potential to transform into a
divergenceless spinor current. Indeed, in $1+1$
dimensions, it is algebraically possible for $N=1$ supersymmetry to include
central charges, so this case is expected to be different.
In fact, explicit examples are constructed in \OW, demonstrating that
$1+1$ dimensional models need not have extended supersymmetry in the
presence of topological charges and that the topological charges
themselves need not appear as the central charges even in the $N=1$
supersymmetric case.
Consequently, we find
the distinction that our argument naturally draws between $1+1$ and
higher dimensions especially compelling, as this reflects
exactly the evidence accumulated from the studies of numerous
specific models, as well as what is known from the general
study of supersymmetry algebras.

\newsec{\B Equations and Inequalities}

We have now established that (in $2+1$ or more spacetime dimensions)
$N=1$ supersymmetric theories with a
topological charge automatically have $N=2$ supersymmetry in
which the topological charge appears as the central charge.
What does this imply for the properties of topologically non-trivial
field configurations? We will take up the cases of solitons and
instantons in turn.

\subsec{The Case of Solitons}

To begin our consideration of \B relationships for solitons, we note
that the result of the preceding section implies that
supersymmetric theories with topological
solitons necessarily exhibit \B bounds and equations \OW,
as a consequence of the representation theory of extended
superalgebras \ref\representation{A. Salam and J. Strathdee,
{\it Nucl. Phys.} {\bf B80} (1974) 499.}.
In such an extended superalgebra, the energy
of a field configuration is bounded from below by the magnitude
of its central
charge \ref\susybks{S. J. Gates, M.T. Grisaru, M. Ro\v cek, and
W. Siegel,
{\it Superspace, or One Thousand and One Lessons in
Supersymmetry}
(Benjamin/Cummings, Reading MA, 1983)\semi
J. Wess and J. Bagger, {\it Supersymmetry and
Supergravity} (Princeton University Press, Princeton, NJ, 1983).},
in this case the topological charge. This is the \B
bound. Furthermore, a field configuration which saturates such
a bound is necessarily annihilated by a linear combination
of supercharges. This algebraic condition
serves the role of the \B equation.
Since supercharges can generically be
represented by first order differential operators (first-order
in spacetime, non-linear in field space), this means that field
configurations that saturate a \B bound in a supersymmetric
theory satisfy a \B equation.\foot{Sometimes the simple application
of the method described in Section 2 for obtaining the second
supercurrent initially will yield a
non-local equation here. This can in general be remedied by adding
appropriate auxiliary fields. Such a use of auxiliary fields is
already familiar from more conventional treatments of solitons even
outside supersymmetric models, as
exemplified by the gauge fields in the $2+1$ dimensional
$\bf CP^1$ model or abelian Chern-Simons model with vortices.}

Second, this in turn implies that {\it any} theory, supersymmetric
or not, with topological solitons exhibits \B relationships \letters
\bogex.
We review the idea briefly here.
Consider a theory with a topological
charge $T_0$ and energy functional of the
theory $E_0$. Now construct a supersymmetric extension of this
theory. Note that any field configuration of the original theory is
automatically a field configuration of the supersymmetric
extension. Furthermore, since a topological charge is conserved
without using the equations of motion, the extended theory has a
topological charge $T_s$ that is {\it  identical} to the topological
charge $T_0$ of the original theory. Finally, it is generally possible
to construct a supersymmetric extension such that the energy of
any field configuration of the original theory is the same whether
evaluated in the original theory (call this energy $E_0$)
or its supersymmetric
extension (call this energy $E_s$) \bogex. (We will have more to say
on this point shortly.)

Now consider a field configuration of the
original theory, and evaluate its energy and topological charge in
both the original and extended theories. We see immediately, then,
that $T_0 = T_s$ and $E_0 = E_s$.
But we know that $E_s \ge |T_s|$ for any field configuration of the
supersymmetric extension, due to the extended superalgebra that
appears in a model. Thus for any field
configuration of the original theory, we have the \B bound
\eqn\bogorig{E_0\ge |T_0|~.}
Furthermore, if a field configuration of the original theory
satisfies $E_0 = |T_0|$, then it also satisfies $E_s = |T_s|$ when
viewed as a field configuration of the supersymmetric extension.
Now any field
configuration with $E_s = |T_s|$ satisfies a \B equation in the
supersymmetric theory. But for the field configuration in question,
this equation involves only the fields of the original theory, and
thus this equation {\it is} the \B equation of the original theory,
satisfied by any field configuration for which $E_0 = |T_0|$.
(Elimination of the extra superpartners or auxiliary fields introduces
no additional derivatives.)
Thus the requirement that a theory with
a topological charge have a consistent supersymmetric extension
implies that a theory with a topological charge --- even a
non-supersymmetric theory --- necessarily exhibits \B relationships.

It is worthwhile to make some comments regarding the construction of
supersymmetric extensions for which $E_0 = E_s$ for any field
configuration of the original theory.  One might be concerned that the
supersymmetric extension of a theory would have new, additional
interactions involving only the fields of the original theory, which
would then invalidate this equality between the two energy functionals
for the field configurations of interest. (Obviously
interactions involving both original and extra fields pose no problem,
as they vanish when the extra fields vanish.) As we discuss below, this
is not a problem.

The reason this is not a problem is that we do not need to find some
minimal supersymmetric extension; any supersymmetric extension, even one
with a large number of additional fields and one that is
non-renormalizable, suffices. Thus, it is only
necessary that there be some supersymmetric extension with the desired
equality between the two energy functionals.

For example, suppose the original theory has a Yukawa coupling. If we
embed the scalar and the fermion in the same superfield, supersymmetry
will require a quartic scalar self-interaction that is not present in
the original theory.  Instead, however, for our purposes, we choose a
supersymmetric extension with two superfields, one of which contains the
original scalar and the other of which contains the original fermion.
This ensures that, for example, the quartic scalar interactions required
by supersymmetry in this case (after elimination of auxiliary fields)
always involve some of the additional
fields not present in the original theory, and so these terms make
no contribution to the energy when the field configuration is a field
configuration of the original theory.

We can merge this approach with a technique introduced in \bogex\ for
constructing appropriate supersymmetric extensions of scalar potentials
to produce a general procedure for obtaining a supersymmetric extension
of the original Lagrangian with a suitable energy functional. The
prescription below depends on certain dimension-independent features of
supersymmetry, such as superfields, supercovariant derivatives, and
auxiliary fields.
It details a systematic and comprehensive (if sometimes cumbersome)
procedure for obtaining an appropriate supersymmetric generalization
that meets our requirements, although often one need not invoke this somewhat
elaborate construction.

Suppose we have a some Lagrangian terms ${\cal L}_{int}$
which we wish to
extend in a suitable supersymmetric fashion.   (As our notation suggests, in
a typical application these would be  interaction terms, although there is
nothing in principle to prevent one from applying this to any subset or
even all of the terms in a given  Lagrangian, and so this notation should not
be taken too literally.)  For each field included in these interactions, we
associate a  superfield.
Then by inserting appropriate
supercovariant derivatives if necessary (so as to project out the
desired Lorentz components of the superfields), one can obtain a
superfield
for each original field in which the original field appears as the
lowest
({\it i.e.}, $\theta=0$) component. We will call this superfield the
associated superfield of a given field. For example, in $3+1$
dimensions, one could place a fermionic field in a scalar superfield
$\Phi$; then $D_\alpha \Phi$ would be the associated superfield, since
it has the fermion as its lowest component. Alternatively, one might
choose as the associated superfield a fermionic superfield, a chiral
superfield endowed with a spinor index, which already has the fermion as
its lowest component.

Note that in this construction, one would generally view a complex field
$S+iP$ in the original theory as two separate real fields. One then
introduces two associated superfields in the extended theory, one for
$S$ and one for $P$. Depending on the dimension, it may be necessary to
take each of these associated superfields to be complex, so that a
single complex field in the original theory is enlarged into two complex
superfields in the extended theory.

Replacing each field in ${\cal L}_{int}$ by its associated superfield,
one obtains a superfield expression ${\cal L}^s_{int}$
which has ${\cal L}_{int}$ as its lowest ({\it i.e.}, $\theta=0$) term.
We also introduce an additional superfield $\Xi$, along with a quadratic
term for $\Xi$, and a term coupling $\Xi$ to $\sqrt{{\cal L}^s_{int}}$
in the superpotential.\foot{The
square
root poses no problems; a Taylor series in Grassman quantities
terminates. If
${\cal L}^s_{int}$ has no terms devoid of fermions, one can simply use
$1+{\cal L}^s_{int}$ instead. Since our work is entirely classical,
renormalizability is of no concern to us.}
The idea is that $\Xi$ has some auxiliary field, call it $B$, so that in
terms of components, the supersymmetric Lagrangian contains both a
$B^2$ term and a $B \sqrt{{\cal L}_{int}}$ term. Eliminating the
auxiliary field $B$, one reproduces ${\cal L}_{int}$ of the original
theory. The other terms in the supersymmetric Lagrangian involve the
other non-auxiliary components of $\Xi$ as well as other components of
the associated superfields, and so vanish when we set the extra physical
fields of the extended theory to zero. Thus when the additional fields
vanish, the
Lagrangian, and hence the energy functional, too, for the supersymmetric
theory reduces to that of the original theory.

Obviously, to implement this, one needs to use superfields appropriate to
the given dimension
\ref\sufld{W. Siegel, {\it Phys. Lett.} {\bf 80B} (1979) 220 \semi
B.E.W. Nilsson, {\it Nucl. Phys.} {\bf B174} (1980) 335\semi
L.B. Litov, Preprint JINR-E2-83-467-mc, July 1983, submitted to
   {\it Bulg. Phys. J.}\semi
X.A. Zhou, {\it Can. J. Phys.} {\bf 66} (1988) 757.}.  However, because
we are not restricted to some particular supersymmetric extension,
it is in fact very easy to formulate a supersymmetric extension with
the desired properties in a dimension independent way.

To begin to understand the flexibility of such a construction, it is
useful first to consider the familiar case of $3+1$ dimensions.  Here,
one can use in our construction $\int d^4\theta \Xi^*\Xi$
as the quadratic term, where $\Xi$ is a chiral superfield which has its
auxiliary field at ${\cal O}(\theta^2)$, and
$\int d^2\theta\sqrt{{\cal L}^s_{int}}\Xi + h.c.$ as the superpotential
term.
This formulation is not unique, however.  We can introduce a potential
superfield $\Psi$, for example,
with the identification $\Xi ={\bar D}^2\Psi$.
The auxiliary field
which appeared at order $\theta^2$ in $\Xi$ appears at order $\theta^4$
in $\Psi$.  The quadratic
term is of the form $\int d^4\theta D^2{\bar\Psi} {\bar D}^2\Psi$,
while the superpotential term is
 $\int d^4\theta \Psi \sqrt{{\cal L}^s_{int}} + h.c$.  Notice that the
field $\Psi$ is not subject to any restrictions (and does not correspond
to an irreducible representation of the supersymmetry algebra).

We now generalize this second formulation to arbitrary dimension.
Suppose we are considering a theory in $d$
spacetime dimensions, in which the superspace has a spinor coordinate
$\theta^\alpha$ with $2s$ components.  If we consider a scalar superfield $\Xi$
in this superspace, its highest component is a scalar field $B$,
which will be the auxiliary field we will use.  The necessary quadratic
term is $\int d^{2s}\theta D^s\Xi D^s\Xi$, while the superpotential term
that generates the interaction is
$\int d^{2s}\theta \sqrt{{\cal L}^s_{int}}\Xi$.
This generates the necessary Lagrangian terms.  Additional terms are
also generated, but these all vanish when the other fields in $\Xi$ (all of
which appear dynamically in these superspace
action terms)
are set to zero.\foot{Of course, the superfield written down here need not
correspond to an irreducible supersymmetry representation.  This is not
a problem for our construction,
although if one wishes, one can project out the irreducible
representation that includes the auxiliary field $B$.}  In this way,
then, we see that it is always possible to construct a supersymmetric
generalization of a given field theory, with the property that when
the extra physical fields added vanish, the energy functional of
the extended theory reduces to that of the original theory.

In practice, one need not necessarily invoke this construction,
even in cases in which
the minimal supersymmetric extension might seem to impose additional
and unacceptable relations among the component fields. For example, consider
a non-abelian gauge theory with a scalar field, say in the adjoint
representation, as occurs in monopole theories. Coupling a scalar
superfield to a vector superfield in a minimal way does produce
additional scalar potential terms (the so-called $D$-terms), but it
turns out that these
automatically vanish when one sets the extra component fields to zero
\bogex.
Thus, the reader should remember that on the one hand, we have obtained
a general construction that demonstrates that is is always possible
to define a suitable supersymmetric extension, but that on the other
hand, many times a minimal supersymmetric extension will be satisfactory
for our purposes.

\subsec{The Case of Instantons}

We now move to a consideration of instantons, demonstrating
that if topological
solitons generically exhibit \B relationships, then instantons,
too, generically exhibit action bounds, as well as ``self-duality''
equations when these action bounds are saturated.

Consider a $d$-dimensional theory with topological instantons.
This means
that the Euclidean theory, with action $S_d$, has field configurations with
non-zero instanton number $I_d$. The instanton number reflects the topological
classification of
the finite action field configurations over $d$-dimensional Euclidean
space. Field configurations of
different instanton number are in different topological sectors.

Suppose we enhance this $d$-dimensional theory by
adding a time coordinate and extra fields as necessary (e.g.,
timelike components for fundamental vector fields) to make
a Lorentz invariant $d+1$-dimensional Minkowskian
theory.\foot{We also adjust an overall sign, so the Euclidean action
is positive semi-definite, while the corresponding terms appear
with the opposite sign in the Minkowski action.} This step
is essential to our argument.  Consequently, if the original
action has dimension-specific expression, such as a term that
can only be written with an epsilon tensor, our argument will not
apply.  However, it is only the appearance of such an expression in
the action that poses a problem; an epsilon tensor in the definition
of the topological index, for example, poses no obstacle at all.

Now obviously, the instanton number becomes a conserved topological charge
of the
enhanced theory. The topological charge arises from the topological
classification of finite energy field configurations over the $d$
dimensional Euclidean space
within the $d+1$ dimensional Minkowski spacetime. Thus the topological
classification of field configurations in the Euclidean theory becomes a
topological  classification in the $d+1$-dimensional theory, in which the
added field components play no role. (Only the fields present in the original
theory appear in the definition of the topological charge.)
Since this classification
of field configurations is topological, under time  evolution, the topological
classification of the field configuration in  space  cannot change,
independent of the equations of motion, and so the  topological
classification that led to an instanton number in $d$ dimensions leads  to a
topological conserved charge in $d+1$ dimensions. As we have argued above,
then, this  $d+1$ dimensional theory necessarily exhibits a \B bound.

Furthermore, in the dimensional enhancement, we need introduce no
interactions different  from the ones in the original theory, although those
interactions may  involve larger fields (e.g., four-component vs.
two-component spinors).  Consequently,  when these additional field components
are set to zero, the enhanced  theory is of the
same form as the  original theory, except
that all the fields have a time dependence as  well. In  the case that these
fields are static, then, the energy $E_{d+1}$ of the  enhanced theory and the
Euclidean action of the original theory $S_d$  are  identical.

Now consider an instanton configuration of the original theory. Clearly,  we
can also interpret it as a field configuration of the enhanced theory in
which  the additional fields all vanish and in which the non-zero fields are
static.  For such a field configuration, the topological charge $T_{d+1}$ in
the  enhanced theory is identically the instanton number $I_d$ in the
original  theory. Such a field configuration interpreted in the $d+1$
dimensional  theory has non-zero topological charge, and as such, cannot
violate the \B bound of that theory. Hence, $E_{d+1}\ge|T_{d+1}|$. But, since
for these field configurations, $E_{d+1}=S_d$ and $T_{d+1}=I_d$, it follows
that \eqn\actionbd{S_d\ge |I_d|~.}

If the instanton saturates the bound \actionbd, then not only is
$S_d=|I_d|$, but, interpreted in the enhanced theory
$E_{d+1}=|T_{d+1}|$.
For field configurations for which $E_{d+1}=|T_{d+1}|$, we know that our
previous analysis produces a \B equation that this configuration must  satsify.
Since this configuration is necessarily static, this \B equation which  is
satisfied in $d+1$ spacetime dimensions is actually an equation in $d$
dimensional space. This equation is an equation in $d$ dimensional  Euclidean
space satisfied by any instanton configuration that saturates the action bound
\actionbd.  In other words, it is the ``self-duality'' equation  for
saturating instantons.

Thus, whenver a theory with instantons can be enlarged to Minkowski
theory in one higher dimensions -- which is the typical case -- those
instantons will exhibit \B relationships.  In fact, even in a theory
with a dimension-dependent term, one might still be able to employ our
results.
Consider a Lagrangian which contains a set of terms that can be dimensionally
enhanced, plus a single term which cannot be.  One first obtains the
\B relationships for the theory without this additional problematic
term using our general methods.  Then one can focus specifically on the
consequences of the additional term to see how it modifies the \B
relationships already obtained.

Note by the way that we have not claimed that any soliton which solves  the
equations of motion in $d+1$ dimensions becomes, upon dimensional  reduction,
an instanton which solves $d$ dimensional equations of motion. An  obvious
counterexample is provided by a time-dependent soliton of the $d+1$
dimensional theory. Importantly, here we are working in the other  direction,
and an instanton of the $d$ dimensional theory is always a field
configuration  of the enhanced $d+1$ dimensional theory that we have
defined.

\newsec{Closing Comments}

In this paper, we have significantly improved on our earlier analysis  of
topological solitons. We have extended our model-independent approach to
include instantons as well as solitons, and to include solitons and
instantons  in $2+1$ or more and two or more dimensions, respectively.  Most
instructively, our analysis of the appearance of extended supersymmetry in
models with solitons clearly demonstrated the difference between  topological
conservation laws in $1+1$ and
higher dimensional supersymmetric theories.  The
very  basic and essential way that our analysis breaks down below $2+1$
dimensions -- just as the theoretical evidence requires -- we find as  telling
evidence that our explanation for the appearance of extended  superalgebras in
theories with solitons is based on the truly fundamental aspects of the
problem.

In closing, we remark that we have used supersymmetry in quite a powerful way
to understand
non-supersymmetric field theories. Moreover, our methods are constructive, not
just formal, and give, to our knowledge, the only model-independent approach
to the \B-type properties of solitons and instantons. We note
that this use of supersymmetry gives a systematic approach to index theorems
in instanton and soliton backgrounds, and to demonstrating equivalences between
the bosonic and fermionic excitation spectra in such backgrounds. We are
currently investigating these matters.

One refinement of our results would be the identification of a way to use
supersymmetry directly to obtain results
about instantons, rather than by going through the process
of dimensional enhancement. We imagine that this is possible,
but so far have been unsuccessful in our attempts to do so.

\bigbreak\bigskip\bigskip\centerline{{\bf Acknowledgments}}\nobreak
D.S. acknowledges the support of a Hobart and William Smith
Faculty Research
Grant and of NSF Grant No. PHY-9207859.
D.S. also thanks the LNS Theory Group at
Cornell University, where much of this research was carried out,
for its
hospitality.

\appendix{A}{}

Here we show that, by means of a gauge transformation,
it is always possible to choose the potential
$A$ of \curleqn\ to be divergenceless In the case of
$2+1$ dimensions, this gauge is simply Lorentz gauge.
In the case of $3+1$ dimensions, one can argue this by formal
equivalence to an electrodynamics problem. Let $A^{\mu\nu}$ be a
rank-2 antisymmetric tensor.
Now interpret $j^\nu = -\partial_\mu A^{\mu\nu}$ as an
electromagnetic
current, and let $F^{\mu\nu}$ be
the electromagnetic field generated by this current. This field
strength can, of course, be written as the ``curl'' of some
vector potential.
Thus $A^\prime_{\mu\nu} = A^{\mu\nu}+F^{\mu\nu}$ is
divergenceless,
antisymmetric, and gauge equivalent to $A^{\mu\nu}$.
In higher dimensions, one can make similar
physical arguments showing the existence of such a gauge.

Alternatively, one can use an $i\epsilon$ prescription.
Let $A^\prime = A + \partial \Omega$. (We suppress indices
for simplicity where possible.)
We now show that we can choose $\Omega$ such
that $\partial\cdot A^\prime = 0$. Note first that,
without loss of generality, we
may take $\Omega$ to be divergenceless. One can argue this by
induction.
We know that a vector potential can be gauged to be
divergenceless. Now
consider the gauge transformations of an antisymmetric 2-tensor.
Such
transformations are labeled by gauge equivalence classes of vectors,
and so without loss of generality we can consider only the gauge
transformations given by divergenceless vector potentials. One uses
this as in the argument below to show that one can always gauge
an
antisymmetric 2-tensor to be divergenceless. And now one uses
this result
in considering the gauge transformations of antisymmetric
3-tensors, which
are labeled by gauge equivalence classes of antisymmetric
2-tensors, etc.

Considering only divergenceless $\Omega$,  we see that for
$A^\prime$
to be divergenceless, we must have
\eqn\boxeqn{\partial^2 \Omega^{\mu_2\cdots\mu_{d-2}} =
\partial_{ \mu_1} A^{\mu_1\cdots\mu_{d-2}}~.}
Making an $i\epsilon$ prescription, we can analytically
continue \boxeqn\ to
\eqn\boxe{\Bigl(\partial^2 +i\epsilon\Bigr)
 \Omega^{\mu_2\cdots\mu_{d-2}} =
\partial_{ \mu_1} A^{\mu_1\cdots\mu_{d-2}}~.}
For $\epsilon\ne 0$, $\partial^2 +i\epsilon$ is invertible, so
$\Omega = \lim_{\epsilon\rightarrow 0}(\partial^2 + i\epsilon)^{-
1}\partial A$.
With this choice of $\Omega$, $A^\prime$ is divergenceless.

\listrefs
\bye